%% file: main.tex
\documentclass{scrartcl}
\pdfoutput=1

\input{packages}
\input{preamble}

\title{How to explain ENDF-6 to computers: A formal ENDF format description language}

\author[1]{Georg Schnabel}
\author[2]{Daniel Lopez Aldama}
\author[1]{Roberto Capote}
\affil[1]{NAPC–Nuclear Data Section, International Atomic Energy Agency, Vienna, Austria}
\affil[2]{Centro de Aplicaciones Tecnol\'ogicas y Desarrollo Nuclear, Havana, Cuba}

\begin{document}

\maketitle

\input{abstract}

\pagebreak

\tableofcontents

\input{introduction}
\input{benefits}
\input{formal-language}
\input{language-specification}
\input{language-applications}
\input{summary}

\printbibliography

\pagebreak

\appendix

\input{endf-grammar}
\input{notes-endf-parser}
\input{endf-parserpy-examples}

\end{document}

%% file: packages.tex
\usepackage[sorting=none,url=true,giveninits=true]{biblatex}
\usepackage{listings}
\usepackage{xcolor}
\usepackage{forest}
\usepackage{hyperref}
\usepackage{cleveref}
\usepackage{authblk}
\usepackage{float}
\usepackage{placeins}

%% file: preamble.tex
\newfloat{codesnippet}{thp}{lop}
\floatname{codesnippet}{Listing}
\crefname{codesnippet}{listing}{listings}
\Crefname{codesnippet}{Listing}{Listings}

\definecolor{codegreen}{rgb}{0,0.6,0}
\definecolor{codeblue}{rgb}{0,0,0.6}
\definecolor{codegray}{rgb}{0.8,0.8,0.8}
\definecolor{codepurple}{rgb}{0.58,0,0.82}
\definecolor{backcolour}{rgb}{0.98,0.98,0.96}

\lstdefinestyle{templatestyle}{
    backgroundcolor=\color{backcolour},   
    commentstyle=\color{codegreen},
    keywordstyle=\color{magenta},
    numberstyle=\tiny\color{codegray},
    stringstyle=\color{codepurple},
    basicstyle=\color{codegreen}\ttfamily\footnotesize,
    breakatwhitespace=false,         
    breaklines=true,                 
    captionpos=b,                    
    keepspaces=true,                 
    numbers=left,                    
    numbersep=5pt,                  
    showspaces=false,                
    showstringspaces=false,
    showtabs=false,                  
    tabsize=2
}

\lstdefinestyle{examplestyle}{
    backgroundcolor=\color{backcolour},   
    commentstyle=\color{codegreen},
    keywordstyle=\color{magenta},
    numberstyle=\tiny\color{codegray},
    stringstyle=\color{codepurple},
    basicstyle=\color{codeblue}\ttfamily\footnotesize,
    breakatwhitespace=false,         
    breaklines=true,                 
    captionpos=b,                    
    keepspaces=true,                 
    numbers=left,                    
    numbersep=5pt,                  
    showspaces=false,                
    showstringspaces=false,
    showtabs=false,                  
    tabsize=2
}

\newcommand{\lex}[1]{\lstinline[style=examplestyle]{#1}}
\newcommand{\ltp}[1]{\lstinline[style=templatestyle]{#1}}

%% file: abstract.tex
\begin{abstract}
The ENDF-6 format, widely used worldwide for storing and disseminating nuclear data, is managed by the Cross Sections Evaluation Working Group (CSEWG) and fully documented in the ENDF-6 formats manual. This manual employs a combination of formal and natural language, introducing the possibility of ambiguity in certain parts of the format specification.

To eliminate the possibility of ambiguity, this contribution proposes a generalization of the formal language used in the ENDF-6 formats manual, which is precisely defined in extended Backus-Naur form (EBNF).

This formalization also offers several other advantages, notably reducing the complexity of creating and updating parsing and validation programs for ENDF-6 formatted files.
Moreover, the availability of a formal description enables the automatic mapping of the low-level ENDF-6 representation, provided by a sequence of numbers, to a more human-friendly hierarchical representation with variable names. Demonstrating these advantages, almost the entire ENDF-6 format description has been translated into the proposed formal language, accompanied by a reference implementation of a \mbox{parser/translator}. This implementation leverages the formal format specification to enable reading, writing, and validating ENDF-6 formatted files.
\end{abstract}

%% file: introduction.tex
\section{Introduction}

The Evaluated Nuclear Data File (ENDF) format~\cite{browneditorENDF6FormatsManual2023} stands as a widely embraced standard for sharing and disseminating nuclear data, with its origins dating back to its conception in 1964. Over time, the format has undergone multiple revisions, culminating in the current iteration known as ENDF-6. Managed by the \textit{Cross Section Evaluation Working Group} (CSEWG), the format evolves through a meticulous process involving formal proposals, discussions within CSEWG, and potential adoptions.

Periodic updates to the ENDF-6 format manual~\cite{browneditorENDF6FormatsManual2023} are essential, serving as a reference description for the format. However, the manual's comprehensive nature, blending formal syntax with natural language to describe record associations and higher-level organization, introduces a potential source of ambiguity. While experienced individuals in the nuclear data field can navigate this ambiguity effortlessly, newcomers may find it a barrier to entry.

Another challenge arises in the manual modification of parsing and validation codes to align with updated ENDF-6 specifications, presenting a risk of errors and consuming time from a limited pool of experts. Additionally, there exists a noticeable separation between data producers and consumers in the nuclear data field, partly due to limited ways of interacting with ENDF-6 formatted data.

To address these challenges, the \textit{Generalized Nuclear Data Structure} (GNDS) format has been conceived~\cite{brownSpecificationsGeneralisedNuclear2023,mattoonGeneralizedNuclearData2012}. The motivation stems not only from the need to overcome ENDF-6 limitations but also to enhance the representation of physical processes. While GNDS is envisioned to completely replace ENDF-6 in the future, the current widespread use of ENDF-6 necessitates solutions to improve its accessibility and usability.

In this document, we present an approach that involves the complete formalization of the language used to describe the ENDF-6 format. Leveraging concepts from formal language theory, a subfield of computer science and linguistics, we propose a formal description language that allows for automatic verification and translation of ENDF-6 formatted files into more common representations like JSON, XML, and HDF5. These representations facilitate the interaction and manipulation of nuclear data, which can also be translated back into the ENDF-6 format.
At its core, this work aims to facilitate the creation and sharing of knowledge in the nuclear data domain, aligning it with the research field of knowledge management, e.g.,~\cite{maierKnowledgeManagementSystems2007}.

The subsequent sections of this document delve into the benefits of a formal ENDF format specification language (\cref{sec:benefits}), key concepts related to formal languages (\cref{sec:formal-language}), and the proposed formal description language for ENDF-6 (\cref{sec:formal-endf-format-specification-language}). Additionally, we explore two applications enabled by this formalization (\cref{sec:endf-language-applications}), namely the mapping/translation of ENDF-6 formatted data and format verification, concluding with a summary in \cref{sec:summary}.

It's worth noting that almost the entire ENDF-6 format description in the reference manual~\cite{browneditorENDF6FormatsManual2023} has already been translated into the proposed formal ENDF description language. Furthermore, the implementation of a Python package based on this formalization is available at~\url{https://github.com/IAEA-NDS/endf-parserpy}, offering solutions to many of the challenges discussed in this introduction. Illustrative examples of how to use this package are provided in~\cref{sec:example-use-case-endf-parserpy}.

This package is a complement to other pertinent codes and packages for working with ENDF-6 formatted data, such as \textit{ENDFtk}~\cite{ENDFtk2023,conlinNJOY21NextGeneration2017}, \textit{FUDGE}~\cite{mattoonManagingProcessingNuclear2023,mattoonGeneralizedNuclearData2012}, \textit{DeCE}~\cite{kawanoDeCEENDF6Data2019}, \textit{SANDY}~\cite{fioritoJEFF3CovarianceApplication2019,fioritoNuclearDataUncertainty2017}, and \textit{endf-parser}~\cite{romanoENDFPythonInterface2023}.
More software for nuclear science can be found in the curated list available at~\cite{romanoAwesomeNuclearCurated2023}.

%% file: benefits.tex
\section{Benefits of a formal ENDF format specification language}
\label{sec:benefits}

Before delving into a more detailed discussion on formal languages and their application to describing the ENDF-6 format, let's explore the potential benefits and implications of having a formal ENDF-6 format description. The first three benefits have already been realized, while the latter two represent future possibilities.

\begin{itemize}
\item \textbf{Format verification:}
The formal ENDF-6 format description enables precise verification of whether a given file complies with the ENDF-6 format. A checking program, informed by this formal description, can provide detailed information on any format violations.

\item \textbf{Separation of physical and logical data model:}
The ENDF-6 formats manual defines symbol names and associates them with physical quantities. The formal format description facilitates automatic mapping between the physical data model (storage structure) and logical data model (symbolic representation), allowing users to work with symbol names.

\item \textbf{Translation between data formats:}
Translation between ENDF-6 formatted files and more widely used formats (e.g., JSON, XML, or HDF5) becomes seamless, enhancing inclusivity in both tool development and data access across different domains.

\item \textbf{Automated code generation:}
A formal language describing the ENDF-6 format can serve as the basis for a compiler that generates efficient code in languages like C or C++, automating tasks such as parsing or manipulating ENDF-6 formatted data.

\item \textbf{Automated redundancy detection and elimination:}
The formal format description aids in the elimination of redundant counter variables, streamlining the mapping between logical and physical data models. This automated process enhances efficiency and reduces unnecessary data storage.

\end{itemize}

These benefits collectively aim to make ENDF-6 formatted data more accessible and facilitate validation and manipulation. Improvements in these aspects are anticipated to accelerate the production and consumption of high-quality evaluated nuclear data. As these benefits hinge on a formal language, the next section delves into understanding what a formal language entails.

%% file: formal-language.tex
\section{What is a formal language?}
\label{sec:formal-language}

Formal languages are studied in formal language theory, see e.g.~\cite{mollIntroductionFormalLanguage1988} for an introduction. This field resides at the intersection of computer science and linguistics. For the benefit of nuclear scientists with programming experience, we present essential terms for a better appreciation of the ideas in this paper.
However, knowledge of formal languages is not a prerequisite for understanding the proposed ENDF format description language presented in~\cref{sec:formal-endf-format-specification-language}.
The rigorous definition of the language in terms of a formal grammar is given in~\cref{sec:production-rules-endf-grammar}.

A formal language is precisely defined by a formal grammar, allowing the determination of whether a given sequence of symbols belongs to the associated formal language. In our context, these symbols are letters, digits, and printable special characters, referred to as \textit{terminal symbols}. Formal grammars consist of \textit{production rules}, specifying how \textit{nonterminal symbols} can be replaced by sequences of terminal and nonterminal symbols.

To illustrate, consider a production rule defining the nonterminal symbol $S$ as a sequence of terminal symbols $a$, $b$, $c$:
\begin{lstlisting}
S : "a" "b" "c"
\end{lstlisting}
Terminal symbols are enclosed in quotation marks to distinguish them from nonterminal symbols. With only this production rule, the only valid sequence of nonterminal symbols would be `abc'.

Production rules can include choices, denoted by the $|$ character, as seen in:
\begin{lstlisting}
T : "a" "b" "c" | "d" "e" "f"
\end{lstlisting}
This rule indicates that $T$ can represent either the sequence `abc' or `def'. Nonterminal symbols can also appear on the right-hand side of a production rule:
\begin{lstlisting}
T : S | "d" "e" "f"
\end{lstlisting}
With this rule for $T$ and the one for $S$ introduced earlier, the valid sequences within the associated formal language would be `abc' and `def'.
Because formal grammars typically consist of multiple production rules, a \textit{start symbol} must be specified, serving as starting point for the expansion process. 

Production rules facilitate the \textit{production} of sequences of terminal symbols conforming to the formal grammar. To achieve this, the start symbol is substituted according to the associated production rule, and the process continues until only terminal symbols remain. The resulting sequence of terminal symbols is within the set of syntactically correct sequences defined by the formal language.

Additional notational elements include the question mark, indicating optional presence (e.g., \lstinline{"a" "b"?}), the asterisk denoting zero or more repetitions (e.g., \lstinline{"a" "b"*}), and brackets for grouping subexpressions (e.g., \lstinline{("a" "b" | "c")}). These notational conventions align with the \textit{extended Backus-Naur form}~\cite{zaytsevBNFWasHere2012}.

Parsing, the process of matching nonterminal symbols in production rules to corresponding parts of a given sequence of terminal symbols, is pivotal in many applications. Parsing algorithms, such as Earley~\cite{earleyEfficientContextfreeParsing1970} and LALR~\cite{deremer1969practical,deremerEfficientComputationLALR1982}, are employed for languages that are described by context-free grammars.
Context-free grammars are a complexity class in the \textit{Chomsky hierarchy}~\cite{chomskyAlgebraicTheoryContextFree1963}. The parsing tree, a result of parsing, is a tree-shaped structure associating nonterminal nodes with nonterminal symbols and terminal nodes with terminal symbols.

\begin{figure}[ht]
\centering
\begin{forest}
for tree={rectangle, draw, minimum size=2em, s sep=1.5cm}
[Sentence
[Subject
["ENDF"]
]
["is"]
[Adjective
["great"]
]
]
\end{forest}
\caption{Simple example of a parsing tree.}
\label{fig:simple-parsing-tree}
\end{figure}
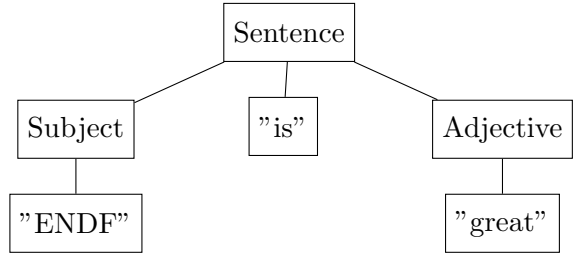

Consider a simple example: parsing the sequence `ENDF is great' with the production rule \lstinline{Sentence : Subject "is" Adjective} results in the parsing tree depicted in \cref{fig:simple-parsing-tree}.

With these foundational concepts introduced, we can now briefly explore the technical perspective of the benefits of having an ENDF-6 format description in a formal language. Parsing a formal ENDF-6 description becomes feasible with readily available parsing algorithms from popular programming language libraries or packages, such as \textit{Lark}~\cite{shinanLarkParsingToolkit2023} for Python. The resulting parsing tree serves as a robust foundation for parsing and translating ENDF-6 formatted files.

Efficient implementation of ENDF-6 parsers and translators involves traversing the nodes of the parsing tree, executing specific actions at each visited node. Further details about the parsing tree derived from a formal ENDF-6 format description and its role in constructing an ENDF-6 parser are provided in~\cref{sec:notes-parser-implementation}.

Basing the design of ENDF parsers on the parsing tree derived from a formal ENDF-6 format description offers advantages. Updates to the ENDF-6 format, described in the formal language, can be seamlessly integrated into an ENDF-6 parser without modifying its code base. Additionally, the tree-centric code design proves robust, as any mistakes in the implementation of actions while visiting nodes are likely to result in program failure, ensuring reliability. This approach also facilitates the creation of robust ENDF-6 format validators.

%% file: language-specification.tex
\section{Formal ENDF format specification language}
\label{sec:formal-endf-format-specification-language}

This section introduces all the elements of the proposed formal ENDF specification language.
While the formal ENDF specification language can be rigorously stated as a list of production rules,
we have decided to take a more colloquial approach to its description, which we believe is more suitable for an audience of nuclear scientists and feels also more natural to the authors anchored in the same domain. 
At the same time, we believe it is straight-forward for computer scientists to formalize the descriptions as a set of production rules.
Moreover, we provide our set of production rules underlying the colloquial descriptions in~\cref{sec:production-rules-endf-grammar}.

Throughout this section and its subsections, we will employ the following color convention.
Strings with templates, meaning that they contain placeholders (also referred to as nonterminal symbols), are printed in a \lstinline[style=templatestyle]{green} font. 
Examples of specific strings in which all placeholders are replaced by terminal symbols that comply with the grammar of the proposed formal ENDF description language are printed in a \lstinline[style=examplestyle]{blue} font.

\subsection{Fundamental elements}
The fundamental elements in the proposed formal ENDF format specification language are \textit{variables}, \textit{arrays} and \textit{numbers}.
These elements appear in (arithmetic) \textit{expressions}, which are important building blocks for higher-level elements of the formal specification language, such as \textit{ENDF records}.

\subsubsection{Variables}
\label{subsubsec:variables}
A variable name can be composed of a combination of digits, letters and underscores but must start with either a letter or an underscore. For instance, \lstinline[style=examplestyle]{AWR}, \lstinline[style=examplestyle]{XMF1} and \lstinline[style=examplestyle]{beta_int} are all valid variable names. 

\subsubsection{Arrays}
\label{subsubsec:arrays}
Arrays are denoted by a variable name followed by one or more comma separated indices enclosed in square brackets. The indices can be given by variable names or integer numbers.  Examples are \lstinline[style=examplestyle]{EN[i]} and \lstinline[style=examplestyle]{EN[5]} to denote one-dimensional arrays, and \lstinline[style=examplestyle]{A[j, k]} and \lstinline[style=examplestyle]{S[q,i,j]} to denote a two- and three-dimensional arrays, respectively.

\subsubsection{Expressions}
\label{subsubsec:expressions}
Expressions can contain integer and decimal numbers as well as variables and arrays. The mathematical operators for addition (\lstinline[style=examplestyle]{+}), subtraction (\lstinline[style=examplestyle]{-}), multiplication (\lstinline[style=examplestyle]{*}) and division (\lstinline[style=examplestyle]{/}) can be present to connect the numbers, variables and arrays. Brackets can also be employed to group subexpressions.
Valid sequences of these symbols must qualify as well-formed arithmetic expressions.
Examples of expressions are \lstinline[style=examplestyle]{(A+B)/C} and \lstinline[style=examplestyle]{A/(B*C)} as well as \lstinline[style=examplestyle]{A+5/(7.0*C)}.
Numbers, variables and arrays alone also qualify as expressions, e.g., \lex{73.4} and \lstinline[style=examplestyle]{A} as well as \lstinline[style=examplestyle]{B[i]} are all valid expressions.
Finally, an example that combines all elements mentioned is given by \lstinline[style=examplestyle]{A+2.3*B[i]/(4-C)}.

\subsection{ENDF records}
\label{subsec:endf-records}

ENDF records are the basic building blocks of an ENDF-6 file.
Various record types exist, each associated with a specific set of numbers stored in adjacent fields within an ENDF-6 formatted file.
The ENDF-6 formats manual~\cite{browneditorENDF6FormatsManual2023} describes the specific order in which the record types are allowed to be assembled in a file.
The manual employs a formal notation for the record specifications, which is closely followed by the formal ENDF specification language proposed in this document.
The following subsections introduce the formal notation for all available ENDF record types, which are the TEXT, CONT, HEAD, END, DIR, TAB1, TAB2, INTG and LIST record. 

\subsubsection{Generic structure of records}
The generic structure of any record type is given by
\begin{lstlisting}
[MAT, MF, MT / ... ] TYPE
\end{lstlisting}
The nonterminal symbols \lstinline{MAT}, \lstinline{MF} and \lstinline{MT} are placeholders for an equally named string or an integer, e.g., \lstinline{MAT} can either stand for \lstinline[style=examplestyle]{MAT} or an integer.
The \lstinline{TYPE} placeholder stands for the string specifying the record type, e.g., \lstinline[style=examplestyle]{HEAD} or \lstinline[style=examplestyle]{CONT}.
A specific example in the proposed formal specification language is
\begin{lstlisting}[style=examplestyle]
[MAT, 1, MT / ... ] TEXT 
\end{lstlisting}
which indicates a TEXT record associated with an MF=1 section.

\subsubsection{TEXT records}
The notation of a TEXT record is given by the template:
\begin{lstlisting}
[MAT, MF, MT/ HL] TEXT
\end{lstlisting}
The placeholder \ltp{HL} can contain a variable~(\cref{subsubsec:variables}) or array~(\cref{subsubsec:arrays}) specification, e.g., \lex{DESCRIPTION[i]}. 
An example of a text record in the proposed format specification language is given by
\begin{lstlisting}[style=examplestyle]
[MAT, 1,451/ DESCRIPTION[i]]TEXT
\end{lstlisting}

\subsubsection{CONT records}
\label{subsubsec:cont-record}
The ENDF-6 manual introduces the following notation for a CONT record:  
\begin{lstlisting}
[MAT,MF,MT/C1,C2,L1,L2,N1,N2]CONT
\end{lstlisting}
The six placeholders from \ltp{C1} to \ltp{N2} stand for expressions~(\cref{subsubsec:expressions}), e.g., \ltp{C1} may be \lstinline[style=examplestyle]{12*NPP}.
The following example demonstrates the the specification of a CONT record in the proposed formal language:
\begin{lstlisting}[style=examplestyle]
[MAT, 14, MT/ EG[k], ES[k], 0, 0, 0, 0] CONT
\end{lstlisting}
Please note that the numbers in an ENDF-6 file corresponding to \ltp{C1} and \ltp{C2} must be floats whereas the numbers corresponding to \ltp{L1}, \ltp{L2}, \ltp{N1} and \ltp{N2} integers.
This restriction of the data type is not enforced at the level of the formal language.

\subsubsection{HEAD records}
The notation of a HEAD record in the formal language is identical to the one of the CONT record (\cref{subsubsec:cont-record}) except for the fact that \lex{CONT} needs to be replaced by \lex{HEAD} at the end of the string.
An example of a HEAD record that complies with the current ENDF-6 format is given by
\begin{lstlisting}[style=examplestyle]
[MAT, 3, MT/ ZA, AWR, 0, 0, 0, 0] HEAD
\end{lstlisting}
According to the manual, the first two slots must always store the variables \lex{ZA} and \lex{AWR}.
However, in the proposed formal language we allow for different choices of variable names in these slots, hence 
\begin{lstlisting}[style=examplestyle]
[MAT, 3, MT/ MaterialCharge , MassParameters, 0, 0, 0, 0] HEAD
\end{lstlisting}
would also be a valid HEAD record specification.

\subsubsection{END records}

The ENDF-6 format makes use of \textit{section end} (SEND), \textit{file end} (FEND), \textit{material end} (MEND) and \textit{tape end} (TEND) records to indicate the end of the respective section and employs the following notations:
\begin{lstlisting}
[MAT,MF, 0/ 0.0, 0.0, 0, 0, 0, 0] SEND
[MAT, 0, 0/ 0.0, 0.0, 0, 0, 0, 0] FEND
[ 0, 0, 0/ 0.0, 0.0, 0, 0, 0, 0] MEND
[ -1, 0, 0/ 0.0, 0.0, 0, 0, 0, 0] TEND
\end{lstlisting}
The manual states that blanks instead of the six zeros after the slash are also valid. 
For the time being, we propose to denote these record types simply as \lex{SEND}, \lex{FEND}, \lex{MEND} and \lex{TEND} in the formal ENDF specification language.

\subsubsection{DIR records}
The template for a DIR record is given by
\begin{lstlisting}
[MAT, MF, MT/ blank, blank, L1, L2, N1, N2] DIR
\end{lstlisting}
The first two slots must contain the string \lex{blank}.
The placeholders \ltp{L1}, \ltp{L2}, \ltp{N1} and \ltp{N2} contain expressions (\cref{subsubsec:expressions}). 
An example for the specification of a DIR record in the proposed formal language is
\begin{lstlisting}[style=examplestyle]
[MAT, 1,451/ blank, blank, MFx[i], MTx[i], NCx[i], MOD[i]]DIR
\end{lstlisting}

\subsubsection{TAB1 records}
\label{subsubsec:tab1-record}
A TAB1 record in the proposed formal description language is of the following form:
\begin{lstlisting}
[MAT,MF,MT/ C1, C2, L1, L2, NR, NP/ XINT / Y]TAB1
\end{lstlisting}
The six placeholders from \ltp{C1} to \ltp{N2} stand for expressions~(\cref{subsubsec:expressions}), e.g., \ltp{L1}  may be \lex{12*(UX-5)}.
The placeholders \ltp{XINT} and \ltp{Y} stand for variable names~(\cref{subsubsec:variables}).
An example of a TAB1 record in the proposed formal language is given by
\begin{lstlisting}[style=examplestyle]
[MAT, 6, MT/ 0.0, mu[j,k], 0, 0, NRP, NEP / Ep / f ]TAB1
\end{lstlisting}

\subsubsection{TAB2 records}
The notation of the TAB2 record is of the form:  
\begin{lstlisting}
[MAT,MF,MT/ C1, C2, L1, L2, NR, NP/ XINT]TAB2
\end{lstlisting}
All the placeholders in this template are also present in the TAB1 record (\cref{subsubsec:tab1-record}) and their description also applies here.
The following example demonstrates the specification of this record type:
\begin{lstlisting}[style=examplestyle]
[MAT, 6, MT/ SPI, 0.0, LIDP, 0, NR, NE / Eint ]TAB2
\end{lstlisting}

\subsubsection{INTG records}

For an INTG record we have the following template:
\begin{lstlisting}
[MAT, MF, MT / II, JJ, KIJ {NDIGIT} ] INTG
\end{lstlisting}
The placeholders \ltp{II}, \ltp{JJ}, \ltp{KIJ} and \ltp{NDIGIT} represent expressions~(\cref{subsubsec:expressions}).
The \ltp{NDIGIT} placeholder does not appear in the notation of an INTG record in the ENDF-6 formats manual but is crucial information to properly interpret the data stored as an INTG record in an ENDF-6 formatted file.
An example of an INTG record in the proposed ENDF format description language is given by
\begin{lstlisting}[style=examplestyle]
[MAT,32,151/ II[k], JJ[k], KIJ[k] {3} ]INTG
\end{lstlisting}

\subsubsection{LIST records}
\label{subsubsec:list-record}

The template for a LIST record is given by
\begin{lstlisting}
[MAT,MF,MT/ C1, C2, L1, L2, NPL, N2/ LIST_BODY ] LIST
\end{lstlisting}
The placeholders from \ltp{C1} to \ltp{N2} stand for expressions~(\cref{subsubsec:expressions}).
The placeholder \ltp{LIST_BODY} represents a comma separated list of one or more \ltp{LIST_BODY_EXPRESSION}. 
A \ltp{LIST_BODY_EXPRESSION} can be either an expression (in the sense of \cref{subsubsec:expressions}) or a \ltp{LIST_BODY_LOOP}. 
The template for the \ltp{LIST_BODY_LOOP} is of the form \ltp{\{ LIST_BODY \}\{COUNTER=START to STOP\}}.

We hightlight that these definitions of placeholders imply that a \ltp{LIST_BODY} can potentially contain itself, so we are dealing with a recursive definition.
This also means that the description of nested loops is possible.
The placeholder \ltp{COUNTER} stands for a variable name (\cref{subsubsec:variables}).
Both placeholders \ltp{START} and \ltp{STOP} represent expressions~(\cref{subsubsec:expressions}).

An example of a \ltp{LIST_BODY_LOOP} is \lex{\{E[i],WE[i]\}\{i=1 to NEI\}}.
Another example of a nested loop specification is given by \lex{\{\{V[m,n]\}\{n=m to MPAR*NRB\}\}\{m=1 to MPAR*NRB\}}.   
As the notation of a LIST record is certainly the most complex case among all record types,
let's consider a couple of examples of LIST records in the proposed formal ENDF specification language:  
\begin{lstlisting}[style=examplestyle]
[MAT, 6, MT/ 0.0, E[j], LANG, 0, NLW[j], NL[j] /
                       {A[j,l]}{l=1 to NLW[j]} ]LIST
\end{lstlisting}
As can be seen, line breaks can be included between elements of the list body for a better visual appearance.
The following example showcases a nested loop in the list body:
\begin{lstlisting}[style=examplestyle]
 [MAT,33,MT/ 0.0, 0.0, LT, LB, 2*NP, NP/
         {Ek[k] , Fk[k]}{k=1 to (NP-LT)}
            {El[k] , Fl[k]}{k=1 to LT} ]LIST
\end{lstlisting}

The formal language to describe LIST records introduced so far is sufficient for a large majority of use cases of this record type in the current ENDF-6 format.
However, for rare cases we need to introduce a further language element. 
The keyword \lex{PADLINE} can be inserted in a \ltp{LIST_BODY} to indicate that the file pointer should be advanced to the next line of the ENDF-6 file before consuming further numbers. 
It can appear at almost any position inside a \ltp{LIST_BODY}.
It is only forbidden within the second bracket pair of \ltp{LIST_BODY_LOOP} where the counter variable and its range are specified. 

An example where the \lex{PADLINE} keyword is necessary in the current ENDF-6 format description is given by
\begin{lstlisting}[style=examplestyle]
[MAT,2,151/ 0.0, 0.0, 0, NRS, 6*NX, NX /
    {ER[n], {GAM[m,n]}{m=1 to NCH} PADLINE}{n=1 to NRS} ]LIST
\end{lstlisting}
The inner \ltp{LIST_BODY_LOOP} states that \lex{NCH} elements of the list body need to be consumed, and hence a parser relying on this specification would advance the file pointer a number of \lex{NCH} fields in an ENDF-6 formatted file.
Because a line break occurs after every six fields, it may happen that the file pointer ends ups at some intermediate position in a line of the ENDF-6 file. 
In the example, the \lex{PADLINE} keyword indicates that the file pointer should be advanced to the next line after all \lex{NCH} fields have been read and before the next run of the inner loop to consume further elements.

\subsection{Meta elements}
\label{subsec:meta-elements}

Most elements within the formal ENDF specification language, as introduced thus far, are also expressed in a formal and consistent language in the ENDF-6 formats manual. However, when describing repetitions of ENDF records or the conditional presence of ENDF records based on variable values, the manual employs various explanation approaches that blend informal language (i.e., natural language) with formal language.

A fundamental concept behind the formal ENDF specification language is to establish a consistent syntax for expressing repetitions and the presence of ENDF records conditional on variable values. This ensures that the entire ENDF-6 format can be formally described, allowing for automatic parsing by computer programs. In the upcoming subsections, we will delve into the description of these language elements—\textit{repetitions} and \textit{conditional blocks}.

Additionally, we will introduce the concept of \textit{sections} to enhance the organization of ENDF variables in a hierarchical manner. These three elements—\textit{repetitions}, \textit{conditional blocks}, and \textit{sections}—are collectively referred to as \textit{meta elements} because they extend beyond the formal ENDF language elements presented in the ENDF-6 manual, specifically ENDF records. They play a crucial role in formalizing the language for describing how ENDF records are combined, assembled, and organized.

\subsubsection{Repetitions}

Parts of ENDF-6 formatted data consist of a specific ENDF record type or a sequence of ENDF records that need to be repeated a certain number of times.
The number of repetitions is usually given by the value of a variable that has been read from an earlier record of the ENDF-6 formatted file.
To formally express repetitions, we propose the following template:
\begin{lstlisting}
for COUNTER=START to STOP:
   ...
endfor
\end{lstlisting}
The placeholder \ltp{COUNTER} needs to replaced by a variable name~(\cref{subsubsec:variables}).
The placeholders \ltp{START} and \ltp{STOP} represent expressions~(\cref{subsubsec:expressions}).
This notation indicates that the statements inside the repetition block, e.g., a sequence of ENDF records~(\cref{subsec:endf-records}), are repeated \ltp{STOP-START+1} times in the ENDF-6 formatted data file.
Importantly, repetitions can also be nested in exactly the same way loops can be nested in imperative programming languages, such as C++ or Python.

An example of a simple repetition in the proposed ENDF format specification language is
\begin{lstlisting}[style=examplestyle]
for i=1 to NWD:
    [MAT, 1,451/ DESCRIPTION[i]]TEXT
endfor
\end{lstlisting}
A more complex (hypothetical) example involving nesting repetitions is given by
\begin{lstlisting}[style=examplestyle]
[MAT, MF, MT/ 0.0, 0.0, 0, 0, NUM, 0] CONT  
for i=5 to NUM+5: 
  [MAT, MF, MT/ one_dim_array[i], 0.0, 0, 0, 0, 0] CONT
  for j=1 to i:
    [MAT, MF, MT/ two_dim_array[i, j], 0.0, 0, 0, 0, 0] CONT 
  endfor
endfor
\end{lstlisting}
This example highlights that array indices do not necessarily need to start at zero or one, which would be a common convention in many programming languages. 
Or the same statement expressed differently, the formal ENDF format specification language does not enforce one- or zero-based array indices.

\subsubsection{Conditional blocks}
The ENDF-6 format relies heavily on variables that act as switches to control, e.g., the representation of data following afterwards. 
Different representations require different variables to store the physics data, which in turn implies a different sequence of ENDF records.
For expressing the presence of a block, containg ENDF records and potentially meta elements, conditional on the values of variables before,
we suggest several templates. 

As a template to express the conditional presence of a block we propose 
\begin{lstlisting}
if LOGICAL_EXPRESSION:
   ...
endif
\end{lstlisting}
We will define \ltp{LOGICAL_EXPRESSION} further below but need to mention here that it is an expression that evaluates to a logical value, i.e., \textit{true} or \textit{false}.

To express the concept of alternatives, we introduce the notation
\begin{lstlisting}
if LOGICAL_EXPRESSION:
  ...
elif LOGICAL_EXPRESSION:
  ...
endif
\end{lstlisting}
More than one block initiated by an \ltp{elif} statement can be present. 

Importantly, it is possible that the \ltp{LOGICAL_EXPRESSION} of all branches evaluate to \textit{false} and none of the conditional blocks is included.
Therefore, we also want to enable the specification of a default block that is included if no other branch applies, for which we propose the following notation:
\begin{lstlisting}
if LOGICAL_EXPRESSION:
  ...
elif LOGICAL_EXPRESSION:
  ...
else:
  ...
endif
\end{lstlisting}
There may be several blocks initiated by an \ltp{elif} clause or none at all. 

The placeholder \ltp{LOGICAL_EXPRESSION} stands for expressions~(\cref{subsubsec:expressions} linked together by comparison operators, 
which are denoted by \lex{>}, \lex{<}, \lex{!=}, \lex{==}, \lex{>=}, \lex{<=} and defined in the same way as in C++ and Python.
An example of a simple \ltp{LOGICAL_EXPRESSION} is \lex{A < B+1}.
Simple logical expressions can be combined to more complex ones by using brackets and the logical operators \lex{or} and \lex{and}.
An example of a more complex \ltp{LOGICAL_EXPRESSION} is \lex{(A < B+1 and B < 10) or B >= 5}.

To conclude the discussion of our proposal to describe the conditional blocks in a formal language, we provide some examples.
A first quite simple example is given by
\begin{lstlisting}[style=examplestyle]
if NK > 1:
  [MAT, 12, MT/ 0.0, 0.0, 0, 0, NR, NP/ Eint / Y] TAB1
endif
\end{lstlisting}
Here is another example making use of the \ltp{elif} statement and also featuring more complex specifications of LIST records~(\cref{subsubsec:list-record}):
\begin{lstlisting}[style=examplestyle]
 if LG == 1:
  [MAT, 12, MT/ ES_NS , 0.0, LP, 0, 2*NT, NT/
                  {ES[i], TP[i]}{i=1 to NT} ]LIST
elif LG == 2:
  [MAT, 12, MT/ ES_NS , 0.0, LP, 0, (LG+1)*NT, NT/
                {ES[i], TP[i], GP[i]}{i=1 to NT} ]LIST
endif
\end{lstlisting}
Finally, the following example features the inclusion of an \ltp{else} statement and a more complex logical expression:
\begin{lstlisting}[style=examplestyle]
if NRO!=0 and (NAPS==0 or NAPS==1):
  [MAT, 2,151/ SPI, 0.0, 0, 0, NLS, 0]CONT
else:
  [MAT, 2,151/ SPI, AP, 0, 0, NLS, 0]CONT
endif
\end{lstlisting}

There is one difficulty in translating parts of the ENDF-6 format description into the proposed formal language concerning conditional blocks.
Sometimes the presence of a record with certain variable assignments depends on the value of a variable in the very same record.
In other words, we do not know if the record with specific variable assignments is present unless we are able to obtain partial information on the values of some variables inside it. 
We address this case by allowing an optional suffix to a \ltp{LOGICAL_EXPRESSION} which is of the form \ltp{[lookahead=NUMBER]}.   
This suffix indicates that some variable names appearing in the logical expression may only be defined within a number of \ltp{NUMBER} records afterwards.
An example of this specification option is given by
\begin{lstlisting}[style=examplestyle]
if NLS==0 [lookahead=1]:
  [MAT, 2,151/ SPI, AP, 0, 0, NLS, 0]CONT
endif
\end{lstlisting}
As can be seen, the record containing the variables \lex{SPI} and \lex{AP} is only supposed to be present if \lex{NLS==0} but \lex{NLS} is defined in the very same record.
The \lex{lookahead} suffix can inform a parsing program relying on the formal ENDF description language about this situation.

\subsubsection{Sections}
\label{subsubsec:sections}
The last important meta element of the formal ENDF format specification language are \textit{sections}.
The purpose of a section is to group a block of statements together under a common name.
A section can be defined using the syntax:
\begin{lstlisting}
(SECTION_NAME)
  ...
(/SECTION_NAME)
\end{lstlisting}
The placeholder \ltp{SECTION_NAME} represents a variable~(\cref{subsubsec:variables}) or an array~(\cref{subsubsec:arrays}).
An example of a section definition in the proposed formal ENDF format specification language is given by
\begin{lstlisting}[style=examplestyle]
(subsection[k])
  [MAT, 8, MT/ ZAP, ELFS, LMF, LFS, 0, 0] CONT
(/subsection[k])
\end{lstlisting}
Importantly, because the block enclosed by the section start and end statements can also contain meta elements, sections can be arbitrarily deeply nested. 

We also introduce two additional types of sections we refer to as \textit{list body sections} and \textit{table body sections}.
A list body section is associated with a LIST record~(\cref{subsubsec:list-record}) and is indicated by adding a suffix to a LIST record specification:
\begin{lstlisting}
[MAT, MF, MT / C1, C2 , L1, L2, NPL, N2 / LIST_BODY] LIST (SECTION_NAME)
\end{lstlisting}
The meaning of this notation is that only the variables included in the \ltp{LIST_BODY} are grouped together under the \ltp{SECTION_NAME}
whereas the variables associated with the placeholders from \ltp{C1} to \ltp{N2} are not included. 

Similarly, a table body section can be defined by suffixing a TAB1 and TAB2 record by a \ltp{SECTION_NAME} enclosed in brackets:
\begin{lstlisting}
[MAT,MF,MT/ C1, C2, L1, L2, NR, NP/ XINT / Y]TAB1 (SECTION_NAME)
\end{lstlisting}
This notation indicates that the variables contained in the placeholders \ltp{XINT} and \ltp{Y} are inside the defined section
whereas the variables associated with the placeholders from \ltp{C1} to \ltp{NP} are outside.

The utility of sections lies in the fact that they can be used to (1) establish a nested hierarchy of groups of variables, and
(2) prevent variable name collisions in a potential mapping from the \textit{physical data model} to a \textit{logical data model}.
What this means is explained in~\cref{subsec:data-model}.

\subsection{Some technical refinements for ENDF-6 files in the wild}

In the real world, many ENDF-6 files do not strictly adhere to the ENDF-6 format. However, these deviations do not necessarily impede the interpretation or subsequent processing of ENDF-6 formatted files for other types of application files. This holds particularly true for two categories of inconsistencies, which we refer to as \textit{inconsistent variable assignments} and \textit{inconsistent number specifications}. The following two subsections will elaborate on these two cases, introducing notational elements that can guide parsers on how to handle these inconsistencies.

\subsubsection{Inconsistent variable assignments}

Some numbers associated with specific symbol names are expected to appear several times at different locations within an ENDF-6 formatted file. For example, consider the following simplified excerpt from the current ENDF-6 format:
\begin{lstlisting}[style=examplestyle]
[MAT, 4, MT/ ZA, AWR, 0, LTT, 0, 0]HEAD
[MAT, 4, MT/ 0.0, AWR, LI, LCT, 0, NM]CONT
\end{lstlisting}
Here, the variable name \lex{AWR} appears in both the first and second ENDF record, indicating that the same number is expected at two different locations in an ENDF-6 formatted file. However, practical ENDF-6 files may violate this requirement, and processing codes may still handle them appropriately.

To address this, we propose introducing a language element that informs a parser about the possibility of such inconsistencies for a specific variable. The parser can then accept an argument from the user to determine how to handle these inconsistencies—whether by terminating parsing with an error message or silently ignoring the issue.

We introduce the convention that a variable name can be suffixed by a question mark (\lex{?}) to indicate the potential for an inconsistency. Using this language element, the above example can be modified as follows:
\begin{lstlisting}[style=examplestyle]
[MAT, 4, MT/ ZA, AWR, 0, LTT, 0, 0]HEAD
[MAT, 4, MT/ 0.0, AWR?, LI, LCT, 0, NM]CONT
\end{lstlisting}

A parser implementing this specification may then ignore the assignment of the number to \lex{AWR} in the second record, inconsistent with the assignment in the first record. Finally, note that the proposed formal ENDF specification language allows for the use of a question mark for a variable or array specification appearing in an expression, as demonstrated by \lex{12*NPP?}.

\subsubsection{Inconsistent number specifications}

In addition to inconsistent variable assignments, another common type of inconsistency found in practice involves the storage of numbers in an ENDF-6 file that differ from the numbers expected by the ENDF-6 format. This issue typically arises in fields that, according to the ENDF-6 format, must be zero but are repurposed by evaluators to store auxiliary information.

For example, MT subsections from 51 to 90 in the MF3 section store cross sections for inelastic scattering at excited levels. The ENDF-6 format expects these sections to start with the record:

\begin{lstlisting}[style=examplestyle]
[MAT, 3, MT/ ZA, AWR, 0, 0, 0, 0] HEAD
\end{lstlisting}

In practice, evaluators may use one of the fields that should be zero to store the level number of the excited states. This deviation from the ENDF-6 format is usually not problematic for processing codes since they typically ignore those fields anyway.

To address this inconsistency, we introduce the question mark (\lex{?}) to indicate the potential presence of such inconsistencies in an ENDF-6 file. Depending on a user-defined option, parsers can be directed to silently ignore these inconsistencies or fail with an error message.

The following example demonstrates the application of this language element:
\begin{lstlisting}[style=examplestyle]
[MAT, 3, MT/ ZA, AWR, 0, 0?, 0, 0] HEAD
\end{lstlisting}

As a final remark, a parser may choose by default to ignore any violation due to a non-zero number in an ENDF-6 file that is supposed to be zero according to the ENDF-6 format description.
Many ENDF-6 files in library projects have violations of this type but are, apart from that, correctly formatted.

%% file: language-applications.tex
\section{Applications of the formal ENDF specification language}
\label{sec:endf-language-applications}
Having established the entire formal ENDF format specification language in~\cref{sec:formal-endf-format-specification-language}, we want to revisit two of the applications listed in~\cref{sec:benefits} and discuss them in more technical detail. The first application involves mapping the physical data model into the logical data model, offering the benefit that individuals and downstream programs (e.g., processing codes) can work with semantically meaningful variable names rather than positions in an ENDF-6 file. The second application pertains to the rigorous checking of whether a given ENDF-6 file complies with a formal description of the ENDF-6 format.

\subsection{Mapping the physical to the logical data model}
\label{subsec:data-model}
In this section, we sketch how the formal description of the ENDF-6 format could be harnessed to create a mapping from the \textit{physical data model} to a \textit{logical data model}.
The physical data model defines the organization of information in a data file at the low level, i.e., what data is stored where.
The logical data model defines the representation at a high level, i.e., how quantities (referred to by symbol names) are related to each other and grouped together.

Importantly, the purpose of this section is to develop a basic intuition how this mapping can be established rather than a comprehensive description of a one-to-one mapping between a physical and logical data model. 
Even though such a comprehensive mapping has been implemented in the \textit{endf-parserpy} Python package~\cite{schnabelEndfparserpy2023}, its extensive documentation is outside the scope of this paper and left as future work. 
Please also note that a proper understanding of this section requires a basic knowledge of programming in general and the Python programming language in particular.

\begin{figure}[b!]
\centering
\begin{forest}
  for tree={rectangle, draw, minimum size=2em, s sep=0.5cm}
  [ENDF Data
    [Cross Sections
      [elastic scattering]
      [inelastic scattering]
      [...]
    ]
    [Angular distributions
      [...]
    ]
    [Energy Spectra
      [...]
    ]
  ]
\end{forest}
\caption{Coarse skeleton of the logical data model}
\label{fig:coarse-logical-data-model}
\end{figure}
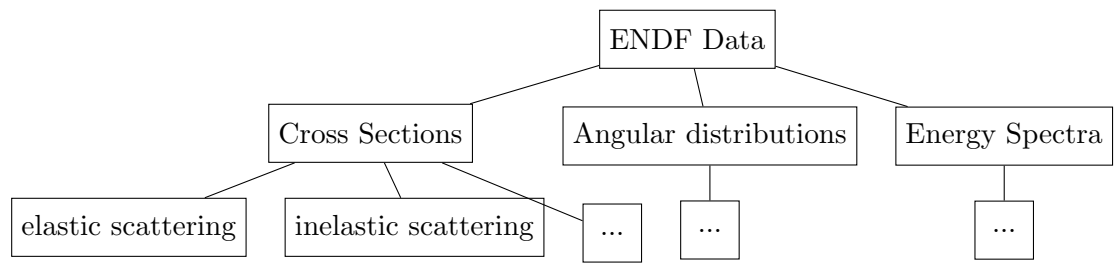

We start out by noting that the data in an ENDF file is already split into distinct groups due to the fact that each line in an ENDF file contains a \textit{MAT}, \textit{MF} and \textit{MT} number. 
The MAT number denotes the material for which the data is given.
The MF number indicates the type of data, such as cross sections or thermal neutron scattering law data.   
The MT number specifies for which reaction the data is stored, e.g., elastic scattering.
More technical details on the categories established by these three numbers can be found in the ENDF-6 formats manual~\cite{browneditorENDF6FormatsManual2023}.
Considering these three numbers, we already obtain a coarse skeleton for the logical data model derived from this hierarchy.
An example of a simple skeleton (skipping the material layer) is depicted in~\cref{fig:coarse-logical-data-model}. 

Taking the programming language Python~\cite{PythonSoftwareFoundation2023} as an example, this hierarchical data structure can be implemented with \textit{dictionaries}, which enable the storage of objects (e.g., numbers, lists, and dictionaries) and provide access to them via user-specified names.
Dictionaries are an implementation of a data structure that is commonly referred to as \textit{hash map} in computer science.
The initialization of such a hierarchical data structure can for instance be achieved by the instructions given in~\cref{fig:basic-endf6struc-init}.
\begin{codesnippet}[tb]
\centering
\begin{lstlisting}[language=Python]
endf_data = dict()
# initialize layer for quantity type
endf_data["cross section"] = dict() 
endf_data["angular distribution"] = dict()
...
# initialize layer for reaction
endf_data["cross section"]["elastic scattering"] = dict() 
...
\end{lstlisting}
\caption{Basic initialization of a hierarchical data structure for ENDF-6 data.}
\label{fig:basic-endf6struc-init}
\end{codesnippet}

The organization of the data below the reaction type level (elastics scattering, inelastic scattering, etc.) will be different depending on the specific quantity type (e.g., cross section) they are associated with.
The organization will also be different depending on the values of specific flags in the ENDF-6 file and the sections introduced in the formal ENDF-6 format specification.   
For a better understanding of the previous sentence, consider the formal ENDF-6 specification of an MF=13 section (photon production cross sections) given in~\cref{fig:endf-recipe-mf13}.
In the following discussion, we refer to this format specification example simply as \textit{ENDF-6 recipe}. 
\begin{codesnippet}[b!]
\centering
\begin{lstlisting}[style=examplestyle]
[MAT, 13, MT/ ZA, AWR, 0, 0, NK, 0]HEAD
if NK>1:
    [MAT, 13, MT/ 0.0, 0.0, 0, 0, NR, NP/ E / sigma_tot ] TAB1
endif
for k=1 to NK:
(subsection[k])
    [MAT, 13, MT/ EG , ES , LP, LF, NR, NP/ E / sigma ] TAB1
(/subsection[k])
endfor
SEND
\end{lstlisting}
\caption{Formal ENDF-6 specification for MF=13 (photon production cross section)}
\label{fig:endf-recipe-mf13}
\end{codesnippet}

All data associated with the variables appearing in the ENDF-6 recipe can be stored inside \lstinline[language=Python]{endf_data["photon production cross section"]}.
Further, let's assume we consider photon production associated with inelastic scattering so we can initialize the respective part of the hierarchical data structure by 
\begin{lstlisting}[language=Python]
endf_data["photon production cross section"] = dict()
prod_xs_dict = endf_data["photon production cross section"]
prod_xs_dict["inelastic scattering"] = dict()
cursec = prod_xs_dict["inelastic scattering"]
\end{lstlisting}

Now we are ready to discuss the mapping of the numbers in an ENDF-6 file into the hierarchical data structure.
The values in an ENDF-6 file associated with variables outside any \textit{section} can be directly mapped into the dictionary, e.g.
\begin{lstlisting}[language=Python]
cursec["ZA"] = corresponding_ZA_value_from_endf_file
cursec["AWR"] = corresponding_AWR_value_from_endf_file
cursec["NK"] = corresponding_NK_value_from_endf_file
\end{lstlisting}

Variables only appearing in a conditional block are only included if the logical expression in its head evaluates to true.
Considering the given ENDF-6 recipe, the variable \lex{sigma_tot} will therefore only be availabe if \lex{NK > 1} so we may continue incrementally building up the data structure by writing
\begin{lstlisting}[language=Python]
if cursec['NK'] > 1:
    cursec["sigma_tot"] = corresponding_list_of_numbers_in_endf_file 
\end{lstlisting}

Finally, let us briefly discuss how the presence of sections can be used for the mapping of numerical data in an ENDF-6 file into a hierarchical data structure. 
In the ENDF-6 recipe, we have a number of \lex{NK} subsections which are distinguished by an index.
The initialization of these subsections in the data structure can for instance be achieved by:
\begin{lstlisting}[language=Python]
cursec["subsection"] = dict()
for k in range(1, cursec["NK"]+1):
    cursec["subsection"][k] = dict() 
\end{lstlisting} 
All the numbers in an ENDF-6 file associated with the variables appearing in a specific section (e.g., \lex{subsection[2]}) would be mapped into the corresponding part in the logical data structure:
\begin{lstlisting}[language=Python]
cursec["subsection"][2]["EG"] = corresponding_EG_value_in_endf_file
\end{lstlisting}  

We conclude this brief discussion of how data can be mapped from the physical data model into the logical data model here.
Evidently, there are several details that have been skipped for the sake of brevity. 
For instance, the visibility of variables defined at different levels in the hierarchy established by sections requires some thought.
We also did not discuss how the ENDF-6 recipe can \textit{actually} be interpreted by a computer program for the purpose of mapping data from an ENDF-6 file into a logical data structure.
Notes on how this can be accomplished are provided in~\cref{sec:notes-parser-implementation}.

\subsection{Basic consistency checking}
\label{subsec:consistency-checking}
Another application of the proposed formal ENDF format specification language is basic consistency checking of ENDF-6 files.

The most basic check concerns the data types of the slots in the various ENDF record types.
For each record type, the data types of all slots are well defined.
A parser can walk through the sequence of ENDF records as specified in the formal ENDF-6 format description and read the associated fields from the ENDF-6 file. 
If, for instance, the parser attempts to read a CONT record~(\cref{subsubsec:cont-record}) and encounters a float in one of the integer slots (\ltp{N1}, \ltp{N2}, \ltp{L1} or \ltp{L2}), it can precisely identify the location of the data type inconsistency in the file and correlate it with the exact line in the formal ENDF-6 description.
This detailed information is certainly helpful for evaluators or any other person that wants to be sure that an ENDF-6 file fully complies with the ENDF-6 format.

A person skimming through the ENDF-6 format manual will also notice that there are sometimes consistency requirements between different fields, which are described by arithmetic expressions.
For instance, consider the following record specification, which is part of the formal description of an MT section inside an MF=12 section:
\begin{lstlisting}[style=examplestyle]
[MAT, 12, MT/ ES_NS , 0.0, LP, 0, (LG+1)*NT, NT/
			  {ES[i], TP[i], GP[i]}{i=1 to NT} ]LIST
\end{lstlisting}
The value read from an ENDF-6 file, corresponding to the second-to-last field, must be equal to the result of the arithmetic expression~\lex{(LG+1)*NT}.
Here, the value associated with the variable~\lex{LG} is obtained from an earlier record, and the value associated with~\lex{NT} is stored immediately afterward in the ENDF-6 file.

Permissible arithmetic expressions, as defined by production rules (refer to~\cref{sec:formal-language}), are an integral part of the system defining the complete formal ENDF format specification language. Parsing and evaluating these expressions can leverage the same programming techniques employed for parsing the full formal ENDF-6 format description. Additional insights into the implementation of this parsing process are offered in~\cref{sec:notes-parser-implementation}.

In summary, the formal ENDF format specification language is a very useful resource for rigorous checking of ENDF-6 files at the level of syntax.
The development of more advanced verification programs that also take into account constraints established by physics can be accelerated by relying on a logical data model.
A brief discussion on the mapping from the physical to the logical data model was provided in~\cref{subsec:data-model}.

%% file: summary.tex
\section{Summary and conclusions}
\label{sec:summary}

We generalized the formal language used in the ENDF-6 formats manual to enable a precise description of the entire ENDF-6 format, eliminating the possibility of ambiguity.
The proposed formal language is defined by a context-free formal grammar, consisting of production rules given in extended Backus-Naur form~(EBNF).

Based on the formal grammar, readily available libraries can be utilized for parsing a formal ENDF-6 description and generating an associated parsing tree.
Programs can leverage this parsing tree for the purpose of ENDF-6 file format verification and providing more user-friendly access to the data of such files. 
Importantly, the amount of work required to update these programs is significantly reduced, sometimes completely eliminating the need to update their code base. 
Furthermore, programs for parsing, verifying and translating ENDF-6 formatted files that rely on the parsing tree can be designed more robustly, reducing the probability of implementation errors. 

The capability of automatically translating ENDF-6 formatted data into more widely used representations and formats, such as JSON, opens the production and consumption of evaluated nuclear data to a wider community.
In particular, modifications in these alternative representations can be translated back to the ENDF-6 format, hence anyone understanding the physical quantities defined in the ENDF-6 formats manual can become a producer of evaluated nuclear data.

The benefits of the availability of a description of the ENDF-6 format in a formal language have already been reaped in practice.
A Python package has been developed that is able to read, write, verify and translate ENDF-6 files by leveraging a formal ENDF-6 format description.
It is foreseen that this package and potentially other packages using a formal ENDF-6 description will facilitate the consumption and generation of high-quality nuclear data in the future.

%% file: endf-grammar.tex
\section{Production rules of the ENDF grammar}
\label{sec:production-rules-endf-grammar}

We provided a description of the formal ENDF specification language in~\cref{sec:formal-endf-format-specification-language} using natural language, which we believe is easier to follow by persons anchored in the domain of nuclear science.
For the sake of rigour and completeness, we include here the production rules in full detail that define the formal grammar underlying the formal ENDF format specification language. 
The notation follows closely the \textit{extended Backus-Naur form} in the notational style employed by \textit{Lark}~\cite{shinanLarkParsingToolkit2023}, which is a Python package for parsing context-free grammars.
\Cref{sec:formal-language} introduced the basic notational elements to specify production rules. 
Consult the documentation of the \textit{Lark} package for an explanation of the  notational elements that were not described in this section.
The start symbol of this grammar is \ltp{endf_recipe}.
At the time of writing, this formal grammar can also be accessed at the url \url{https://raw.githubusercontent.com/IAEA-NDS/endf-parserpy/11c6c8b566fefb9201695391d2c4cbb633510a56/endf_parserpy/endf_lark.py}. 

\begin{lstlisting}
%import common.DIGIT
%import common.LETTER
%import common.NEWLINE
%import common.STRING
%import common.CNAME
%import common.INT
%import common.NUMBER
%import common.ESCAPED_STRING
%ignore " "

endf_recipe : (code_token | NEWLINE)*
code_token: endf_line | for_loop | if_clause | section
endf_line : (list_line | head_or_cont_line | tab1_line | tab2_line
            | text_line | dir_line | intg_line | send_line
            | dummy_line | stop_line | COMMENT_LINE) NEWLINE

// section to define namespace for variables
section: section_head section_body section_tail
section_head : "(" extvarname ")" NEWLINE
section_body : (code_token | NEWLINE)*
section_tail : "(/" extvarname ")" NEWLINE

// control numbers
ctrl_spec : MAT_SPEC "," MF_SPEC "," MT_SPEC
MAT_SPEC : "MAT" | INT
MF_SPEC :  "MF" | INT
MT_SPEC : "MT" | INT

// generic record body
record_fields : expr "," expr "," expr "," expr "," expr "," expr

// comment line
COMMENT_LINE : "#" /.*/

// stop instruction to quit parsing
stop_line : "stop" "(" escaped_stop_message? ")"
escaped_stop_message : "\"" STOP_MESSAGE "\""
STOP_MESSAGE : /[^"]+/

// DUMMY record (read but not processed)
dummy_line : "[" ctrl_spec "/" dummy_body "]" "DUMMY"
dummy_body : (LETTER | DIGIT | " " | ",")

// TEXT record
text_line : "[" ctrl_spec "/" text_fields "]" "TEXT"
text_fields : expr

// HEAD and CONT record
head_or_cont_line : "[" ctrl_spec "/" record_fields "]" CONT_SUBTYPE
CONT_SUBTYPE : "CONT" | "HEAD"

// DIR record
dir_line : "[" ctrl_spec "/" "blank" "," "blank" "," dir_fields "]" "DIR"
dir_fields :  expr "," expr "," expr "," expr

// INTG record
intg_line : "[" ctrl_spec "/" intg_fields "{" ndigit_expr "}" "]" "INTG"
intg_fields : expr "," expr "," expr
ndigit_expr : expr

// TAB1 record
tab1_line : "[" ctrl_spec "/" tab1_fields "]" "TAB1" ("(" table_name ")")?
tab1_fields : record_fields "/" tab1_def
tab1_def : extvarname "/" extvarname
table_name : extvarname

// TAB2 record
tab2_line : "[" ctrl_spec "/" tab2_fields "]" "TAB2" ("(" table_name ")")?
tab2_fields : record_fields "/" tab2_def
tab2_def : extvarname

// LIST record
list_line : "[" ctrl_spec "/" record_fields "/"  list_body  "]" "LIST" ("(" list_name ")")?
list_body : (expr | list_loop | LINEPADDING | "," | NEWLINE)+
list_name : extvarname
LINEPADDING : "PADLINE"

// LIST loop
list_loop : "{" list_body "}" "{" list_for_head "}"
list_for_head : VARNAME "=" for_start "to" for_stop

// SEND record
send_line : "SEND"

// FOR loop
for_loop : "for" for_head for_body "endfor" NEWLINE
for_head : VARNAME "=" for_start "to" for_stop ":"
for_body : (code_token | NEWLINE)*
for_start :  expr
for_stop :   expr

// IF statement
if_clause : if_statement elif_statement* else_statement? "endif"
if_statement : "if" if_head (lookahead_option)? ":" if_body
elif_statement : "elif" if_head (lookahead_option)? ":" if_body
else_statement : "else:" if_body

lookahead_option : "[" "lookahead" "=" expr "]"
if_head : disjunction
if_condition : expr IF_RELATION expr
IF_RELATION  : ">" | "<" | "!=" | "==" | ">=" | "<="
IF_AND : "and"
IF_OR : "or"
if_body : (code_token | NEWLINE)*
// the solution in the answer at https://stackoverflow.com/questions/63493679/backus-naur-form-with-boolean-algebra-problem-with-brackets-and-parse-tree
// was helpful and modified to treat conjunction, disjunction and bracketed expressions
disjunction : disjunction "or" conjunction | conjunction
conjunction : conjunction "and" comparison | comparison
comparison : if_condition | "(" disjunction ")"

// arithmetic expression
// adopted from: http://marvin.cs.uidaho.edu/Teaching/CS445/grammar.pdf (3.3)
expr : addition | subtraction | addpart
addpart : multiplication | division | mulpart
mulpart : minusexpr | extvarname | inconsistent_varspec | NUMBER | DESIRED_NUMBER | bracketexpr

multiplication : addpart "*" mulpart
division : addpart "/" mulpart
addition : expr "+" addpart
subtraction : expr "-" addpart

minusexpr: "-" mulpart
bracketexpr : "(" expr ")"

// allowed variable names (including indices)
extvarname : VARNAME ("[" indexquant ("," indexquant)* "]")?
VARNAME : CNAME
INDEXVAR : CNAME
INDEXNUM : NUMBER
indexquant : INDEXVAR | INDEXNUM

// special number symbol indicating that
// a specific number is expected but it may be different in practice
DESIRED_NUMBER : NUMBER "?"

// special variable symbol indicating
// that if the value of the variable in the current record
// is permitted to be inconsistent with a previously read value
inconsistent_varspec : extvarname "?"

// possible field values
CFIELD: CNAME | "0.0"
IFIELD: CNAME | "0"
\end{lstlisting}

%% file: notes-endf-parser.tex
\FloatBarrier

\section{Some notes on the implementation of an ENDF-6 parser}
\label{sec:notes-parser-implementation}

The objective of this document was to introduce a formal ENDF description language.
It was argued that the availability of a formal description of the ENDF-6 format helps in making ENDF-6 formatted data more accessible,
e.g., by mapping a physical data model into a logical data model, see~\cref{subsec:data-model}.
It was explained in~\cref{sec:formal-language} that parsing algorithms, e.g.~\cite{shinanLarkParsingToolkit2023}, are readily available to leverage formal grammars,
as provided in~\cref{sec:production-rules-endf-grammar}, to construct a parsing tree.
The construction of a parsing tree for an ENDF-6 format description given in the formal language introduced in~\cref{sec:formal-endf-format-specification-language} is therefore a solved problem.

However, the utilization of such a parsing tree to parse/interpret an ENDF-6 formatted file requires programming techniques, which are probably less familiar to persons anchored more in nuclear science than computer science.
Hence, we are going to sketch some essential ideas of how the parsing tree can be utilized. The important take-away message is that even though these programming techniques may be unfamiliar, they are not complicated. 

\begin{figure}[bt]
\centering
\begin{forest}
  for tree={rectangle, draw, minimum size=2em, s sep=1.5cm}
  [endf\_recipe
    [code\_token
      [endf\_line
        [head\_or\_cont\_line]
      ]
    ]
    [code\_token
      [endf\_line
        [head\_or\_cont\_line]
      ]
    ]
    [code\_token
      [endf\_line
        [head\_or\_cont\_line]
      ]
    ]
  ]
\end{forest}
\caption{Example parsing tree of an ENDF-6 recipe}
\label{fig:example-mfmt-recipe-tree}
\end{figure}
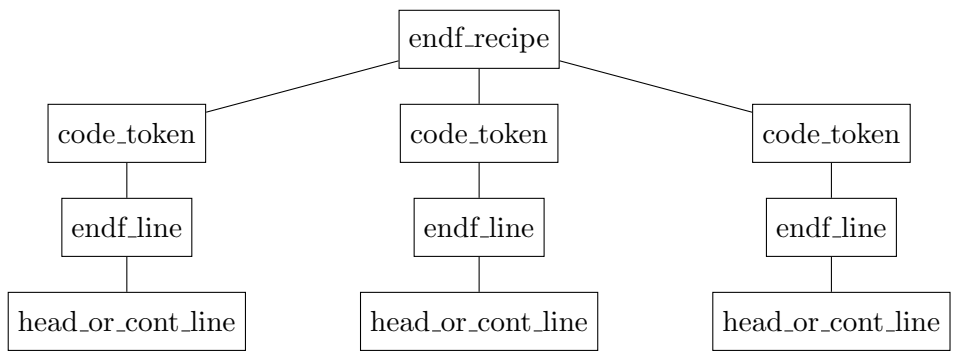

To get started, we restrict ourselves to the level of an MT section.
Consider the following formal description of the header of an MF1/MT451 section:
\begin{lstlisting}[style=examplestyle]
[MAT, 1,451/ ZA, AWR, LRP, LFI, NLIB, NMOD]HEAD
[MAT, 1,451/ ELIS, STA, LIS, LISO, 0, NFOR]CONT
[MAT, 1,451/ AWI, EMAX, LREL, 0, NSUB, NVER]CONT
\end{lstlisting}
The essential production rules in the formal grammar in~\cref{sec:production-rules-endf-grammar} related to this specification are
\begin{lstlisting}
endf_recipe : (code_token | NEWLINE)*
code_token: endf_line | for_loop | if_clause | section
endf_line : (list_line | head_or_cont_line | tab1_line | tab2_line
            | text_line | dir_line | intg_line | send_line
            | dummy_line | stop_line | COMMENT_LINE) NEWLINE
\end{lstlisting}
The start symbol is \ltp{endf_recipe}, which matches a sequence given by \ltp{code_token}'s and/or newline characters.
A \ltp{code_token} can be an \ltp{endf_line} or any of the meta elements described in~\cref{subsec:meta-elements}.
Finally, an \ltp{endf_line} is either an ENDF record~(see \cref{subsec:endf-records}) or any of the elements that were used for debugging ENDF recipes, such as \ltp{dummy_line} or \ltp{stop_line}.
Because the description snippet above contains a HEAD and two CONT records, matched by \ltp{head_or_cont_line}, the generated parsing tree of the formal MF1/MT451 description is of the form depicted in~\cref{fig:example-mfmt-recipe-tree}.
How this parsing tree is represented by a data structure is dependent on the programming language and employed library implementing the parsing. 
For example, in Python it may be represented by nesting tuples and lists, see~\cref{fig:python-tree-init}.
\begin{codesnippet}[t]
\begin{lstlisting}[language=Python]
("endf_recipe", [
    ("code_token", [
        ("endf_line", [
           ("head_or_cont_line", [ ... ]) 
          ]
        )
      ]
    ),
    ...
    ("code_token", [
        ("endf_line", [
           ("head_or_cont_line", [ ... ]) 
          ]
        )
      ]
    )
  ]
)
\end{lstlisting}
\caption{Python code to initialize a tree using tuples and lists}
\label{fig:python-tree-init}
\end{codesnippet}

To interpret an ENDF-6 formatted file, a parser can visit all the nodes in \textit{pre-order traversal},
which means to first visit the current node and then recursively visit in pre-order traversal the child nodes from left to right.
Whenever the parser encounters a special node, such as a \ltp{head_or_cont_line} node, it may employ specialized code to perform some action, e.g., for reading and writing a HEAD or CONT record and advancing the file pointer. 
The Python snippet in~\cref{lst:pre-order-traversal-snippet} exemplifies this type of traversal (assuming that nodes are represented by tuples as in~\cref{fig:python-tree-init}).
\begin{codesnippet}[b!]
\begin{lstlisting}[language=Python]
def traverse_tree(current_node):
    if current_node[0] == "head_or_cont_line":
        # some special code to treat HEAD or CONT records
    else:
        for child_node in current_node[1]:
            traverse_tree(child_node)
\end{lstlisting}
\caption{Code snippet for pre-order traversal of parsing tree}
\label{lst:pre-order-traversal-snippet}
\end{codesnippet}

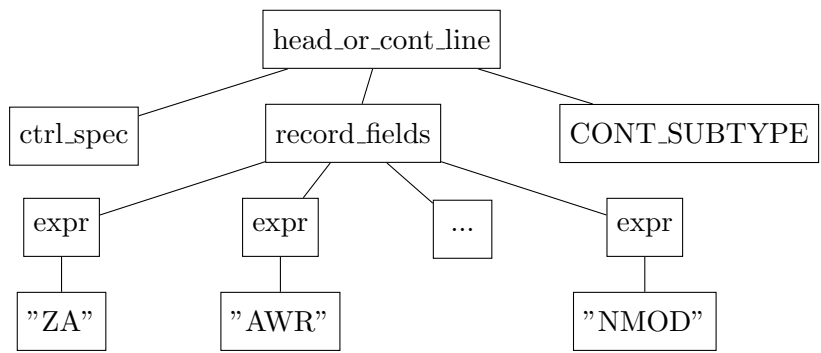
\begin{figure}[b!]
\centering
\begin{forest}
  for tree={rectangle, draw, minimum size=2em, s sep=1.5cm}
  [head\_or\_cont\_line
    [ctrl\_spec]
    [record\_fields
      [expr
        ["ZA"]
      ]
      [expr
        ["AWR"]
      ]
      [...]
      [expr,
        ["NMOD"]
      ]
    ]
    [CONT\_SUBTYPE]
  ]
\end{forest}
\caption{Parsing tree corresponding to a HEAD or CONT record specification}
\label{fig:simple-endf-cont-parsing-tree}
\end{figure}

To discuss the actions an ENDF-6 parser may perform while encountering special nodes,
let's consider the if-branch in~\cref{lst:pre-order-traversal-snippet} that is entered during a visit of a \ltp{head_or_cont_line} node.
The left-most \ltp{head_or_cont_line} node in~\cref{fig:example-mfmt-recipe-tree} corresponds to the first line of the ENDF-6 recipe given above:
\begin{lstlisting}[style=examplestyle]
 [ MAT , 1 ,451/ ZA , AWR , LRP , LFI , NLIB , NMOD ] HEAD
\end{lstlisting}
The essential production rules in the formal grammar in~\cref{sec:production-rules-endf-grammar}, which are utilized to construct the subtree with \ltp{head_or_cont_line} as root are given by
\begin{lstlisting}
head_or_cont_line : "[" ctrl_spec "/" record_fields "]" CONT_SUBTYPE
record_fields : expr "," expr "," expr "," expr "," expr "," expr
CONT_SUBTYPE : "CONT" | "HEAD"
\end{lstlisting}
In this excerpt, we did not include the production rule for \ltp{ctrl_spec} because it would make the display more convoluted without adding any insight on the conceptual level. 
We also excluded the production rule for the nonterminal symbol \ltp{expr} as it stands for an arithmetic expression whose form is more complex and not suited for this brief sketch.
In the adopted convention, nonterminal symbols are written in lower case and terminal symbols in upper case.
The \ltp{head_or_cont_line} node is the root of the subtree shown in \cref{fig:simple-endf-cont-parsing-tree}.

The parser can rely on the form and naming of the production rules, hence it can also rely on the naming of nodes in the parsing tree and make assumptions on possible structures of subtrees.
In the case of an \ltp{head_or_cont_line} node, the parser can proceed by finding the child node named \ltp{record_fields} and then process the associated child nodes named \ltp{expr} one by one to identify the variable names associated with the various slots in a HEAD or CONT record. Simultaneously, it can also read the corresponding numbers from an ENDF-6 file and associate them with the variable names to incrementally map the numbers into a logical data model, which was discussed in~\cref{subsec:data-model}.

\FloatBarrier

\pagebreak
\FloatBarrier

%% file: endf-parserpy-examples.tex
\section{Example use cases of endf-parserpy}
\label{sec:example-use-case-endf-parserpy}

The Python package \textit{endf-parserpy}~\cite{schnabelEndfparserpy2023} enables reading, writing, verifying and translating ENDF-6 formatted data. 
Because this package relies on a formal ENDF-6 format description, it is an example of how the formal description language introduced in~\cref{sec:formal-endf-format-specification-language} can be exploited to facilitate the interaction with ENDF-6 formatted data.
Its design is based on the parsing tree derived from the ENDF-6 format description and the programming techniques outlined in~\cref{sec:notes-parser-implementation}.
Here we demonstrate three possible use cases, which are (1) the manipulation of data in an ENDF-6 file, (2) translation of ENDF-6 to JSON files, and (3) format verification.

The process for manipulating ENDF-6 formatted data follows three steps: reading the data into a nested Python dictionary, manipulating the data, and writing the modified dictionary to an output file. 
To illustrate, assume that we want to modify the \textit{material charge} and \textit{mass parameters}, denoted by the variables \textit{ZA} and \textit{AWR}, respectively, in the section of an ENDF-6 file storing the total cross section (MF3/MT1).
The following Python code snippet implements these three steps:
\begin{lstlisting}[language=Python]
from endf_parserpy import ExtEndfParser
parser = ExtEndfParser()
endf_dict = parser.parsefile("input.endf")
endf_dict[3][1]["ZA"] = 26056 
endf_dict[3][1]["AWR"] = 55.45443
parser.writefile("modified_file.endf", endf_dict)
\end{lstlisting}
This snippet also demonstrates the advantage of working with a logical data model.
Data in the ENDF-6 file can be changed by referring to variable names instead of positions in an ENDF-6 file. 

ENDF-6 sections can also be created from scratch.
Let's assume we want to store a sine wave as total cross section.
We can construct an appropriate dictionary and then output it as an ENDF-6 file:
\begin{lstlisting}[language=Python]
import numpy as np
endf_sec = {
  "MAT": 2925, "MF": 3, "MT": 1,
  "ZA": 29063, "AWR": 62.398,
  "QM": 0.0, "QI": 0.0, "LR": 0
}
xstable = {
  "NBT": [3749], "INT": [2],
  "E": np.linspace(1e-05, 30e6, 1000),
  "xs": np.sin(np.linspace(1e-05, 30e6, 1000))
}
endf_sec["xstable"] = xstable
endf_dict = {3: {1: endf_sec}}
parser.writefile("new_file.endf", endf_dict)
\end{lstlisting}

One may wonder how to determine which fields need to be present in a nested dictionary.
The required structure can be immediately seen from the ENDF-6 description on which the parser relies:
\begin{lstlisting}[style=examplestyle]
[MAT, 3, MT/ ZA, AWR, 0, 0, 0, 0] HEAD
[MAT, 3, MT/ QM, QI, 0, LR, NR, NP / E / xs]TAB1 (xstable)
SEND
\end{lstlisting}
Every variable in this ENDF-6 recipe, which is not enclosed by a section, such as \lex{AWR}, has to be present in the root dictionary.
Because the recipe makes use of a \textit{table body section} (see~\cref{subsubsec:sections}), the variables \lex{NBT}, \lex{INT}, \lex{E} and \lex{xs} must be stored in a subdictionary, which is named \textit{xstable} here.
It may be regarded as a shortcoming that the variables \lex{NBT} and \lex{INT} do not explicitly appear in the formal description.
This choice has been taken to follow closer the notation in the ENDF-6 formats manual~\cite{browneditorENDF6FormatsManual2023}.
On the other hand, the variables \lex{NR} and \lex{NP} are missing in the dictionary.
The parser does not require them because \lex{NR} is inferred from the number of elements in \lex{NBT} (or \lex{INT}) and \lex{NP} from the number of elements in \lex{E} (or \lex{xs}).
They could have also been included in the dictionary but would have been ignored. 

Next, we demonstrate the translation from ENDF-6 files to JSON files.
This translation can be achieved by first reading in the ENDF-6 data into a nested dictionary and then outputting it as a JSON file:
\begin{lstlisting}[language=Python]
import json
endf_dict = parser.parsefile("input.endf")
with open("endf_file.json", "w") as f:
    json.dump(endf_dict, f, indent=2)
\end{lstlisting}

The reverse direction, translating JSON to ENDF-6 files, is also possible:
\begin{lstlisting}[language=Python]
from endf_parserpy.user_tools import sanitize_fieldname_types
with open("endf_file.json", "r") as f:
    endf_dict = json.load(f)
sanitize_fieldname_types(endf_dict)
parser.writefile("endf_out.endf", endf_dict)
\end{lstlisting}
Fieldnames in a JSON file must be strings whereas they are allowed to be numbers in Python.
Calling \lstinline[language=Python]{sanitize_fieldname_types} ensures that fieldnames are converted back to numbers before writing out the content of the dictionary to an ENDF-6 file.
Without this sanitization, the writing operation would fail.

Finally, let's consider format verification.
Being able to successfully read an ENDF-6 file into a Python dictionary means that the source file complies with the ENDF-6 format.
A format violation will lead to failure with a detailed debug output hinting at the specific type of violation.
Here is an example how this output may look like:
\begin{lstlisting}[style=examplestyle]
endf_parserpy.custom_exceptions.ParserException:
Here is the parser record log until failure:

-------- Line 0 -----------
Template:  [ MAT , 4 , MT / ZA , AWR , 0 , LTT , 0 , 0 ] HEAD
Line:     " 1.804000+4 3.961910+1          0          1          0          01837 4  2    1"

-------- Line 1 -----------
Template:  [ MAT , 4 , MT / 0.0 , AWR ? , LI , LCT , 0 , 0 ] CONT
Line:     " 0.000000+0 3.965640+1          0          2          0          01837 4  2    2"

Error message: Expected 39.6191 in the ENDF file but got 39.6564. The value was encountered in a source field named C2
\end{lstlisting}
As seen, the lines of the ENDF-6 file are displayed along with the lines from the ENDF-6 description that are supposed to match them.
In this particular example, the value associated with \lex{AWR} in the second record present in the ENDF-6 file is inconsistent with the value provided in the first record. 
A user may initialize the parser with the argument \lstinline[language=Python]{ignore_varspec_mismatch=True} to parse the file anyway and ignore the inconsistency.